# Pre-scanning Assembly Optimization Criteria for Computed Tomography

Mayank Goswami

Divyadrishti Imaging Laboratory, Department of Physics, IIT Roorkee, Roorkee, India

Department of Physics, IIT Roorkee, Roorkee, India

mayank.goswami@ph.iitr.ac.in

## Abstract
Computerized Tomography assembly and system configuration are optimized for enhanced invertibility in sparse data reconstruction. Assembly generating maximum principal components/condition number of weight matrix is designated as best configuration. The gamma CT system is used for testing. The unoptimized sample location placement with $\pm 7.7\%$ variation results in a maximum 50% root mean square error, 16.5% loss of similarity index, and 40% scattering noise in the reconstructed image relative to the optimized sample location when the proposed criteria are used. The method can help to automate the CT assembly, resulting in relatively artifact-free recovery and reducing the iteration to figure out the best scanning configuration for a given sample size, thus saving time, dosage, and operational cost.

## 1. Introduction

Transmission-based fan beam computerized tomography (CT) scanning operation may require setting the following parameters: (a) voltage/strength of radiation source, (b) sample distance from source, and (c) distance between source and detector or fan beam/cone beam angle. Inter-detector distances and curvature of the scanning array decide the fan beam angle. The parallel beam CT system requires the distance between collimation and, thus, the distance between detectors/inter-detector distances to be set. These CT assembly configuration-related parameters are set to ensure the best transmission of radiation from the source to the detector via sample. The CT systems (especially Gamma CT) have source and detector array assemblies having a fixed distance from each other. The source's voltage and power (in terms of current) must be set so that the radiation beam penetrates through the sample until the detector. A-priori information of sample material may be required to set the value of voltage and power of the source in case of X-ray or ultrasound modality[1]. Operators may, however, test these settings by evaluating the contrast of radiographs for different values. Fan or cone beam angle can be set dynamically for maximum information content[2]. All these parameters require at least one scanning and reconstruction time to clear the doubts.

### 1.1. Motivation

The motivation of this work is to test optimality criteria addressing the points mainly: (a) where to place the object of a given size, (b) what should be the distance between the detector and source, and (c) what should be the distance between detectors in scanning array. These mutually dependent criteria can be integrated into CT system control codes to advise the CT operator for CT assembly optimization prior to the scanning process to provide a relatively better quality of CT reconstruction. The idea is to set CT system components (source, detector, and sample/object) in relative positions to support inverse problem modeling by reducing the scattering noise and sparsity.

## 2. Method and Material

It leaves only source-to-sample distance at the operators' disposal to set. A sub-optimal distance may introduce artifacts such as inactive pixels having zero rays crossing[3]. A typical approach is to ensure that the samples' boundaries are just covered by a radiation beam, as shown in fig. 1. The first arrangement (A1) shows a shaped sample is placed near the detector array (made of two detectors placed symmetrically on the x-axis). The center of the sample is on the x-axis, aligned with the center of the source.

### 2.1. Fan beam CT configuration

The region of interest (ROI) that covers the sample is discretized using a square pixel grid (made of four individual square-shaped pixels) shown in fig. 1(a). The corresponding fixed scanning geometry (referred to as A1) for a single projection is shown in fig. 1(b). The radiation path propagating inside each pixel is shown by a dotted red line originating from the radiation source (red star on the leftmost side) in fig. 1(b). The forward CT algorithm integrates the attenuated radiation counts ($p$) in each pixel along this line based on the respective attenuated coefficient ($\lambda$) and intercept length (L). Figure 1(a) shows these intercept lengths for three different situations (A1, A2, and A3) when the sample is placed on SOD, SOD' and SOD" distances from the source. Similarly, fig. 1(c) shows three situations in 3D when the source and scanning array are rotated 35 times, measuring 36 CT projections/scans symmetrically distributed in 360 degrees. The arrangement of rays for all scans, when overlapped, illustrates a crisscrossed pattern of rays; however, for condition A1, the D>0 is negligible in conditions A2 and A3. The measurement helps to model the CT algorithm into the system of equations solved together. It can be written in the form of equation 1.

### 2.2. Parallel Beam CT configuration

Figure 1(d) shows four different conditions for the parallel beam ultrasound CT scanner. The sensors are housed in a Phased array probe made of 8 sensors, which can be activated or deactivated. At a time, only two sensors are activated and shown by a relatively darker color scheme. The configurations shown in the first two rows represent the first scanning configuration, A1. In this scheme, only the nearest sensors are activated, generating ultrasound beams propagating in the nearest distance (shown in red dotted lines). In any scanning configuration, it is shown that displacing the sample from SOD to SOD" distance between sensor and detectors does not change values of any intercept length L clarifying that changing the sample location cannot help as optimality criteria for parallel beam CT system. The second and fourth rows show the sample (irrespective of whatever distance) is rotated by 90 degrees. Comparing these two situations in A1 and A2, the intercept lengths change. The L1 < L1", L3<L3" and L2"<L2. This explains how changing inter-detector distance from D1 to D2 affects the value of the elements of weight matrix $W$ and thus can be included in the analysis of CT assembly optimization.

Each condition of sample placement thus has a different Matrix (referred to as intercept or weight matrix (W)) of intercept length[4].

$$W_{m \times n}\, \lambda_{n \times 1} = p_{m \times 1} \qquad (1)$$

Where $W_{i,j} = L_{i_{\theta},j}$, $i, j,$ and $\theta$ are the index of the detector, the index of the pixel, and the number of the rotation, $m$ and $n$ are the number of detectors times the number of rotations and the number of pixels used to discretize the sample.

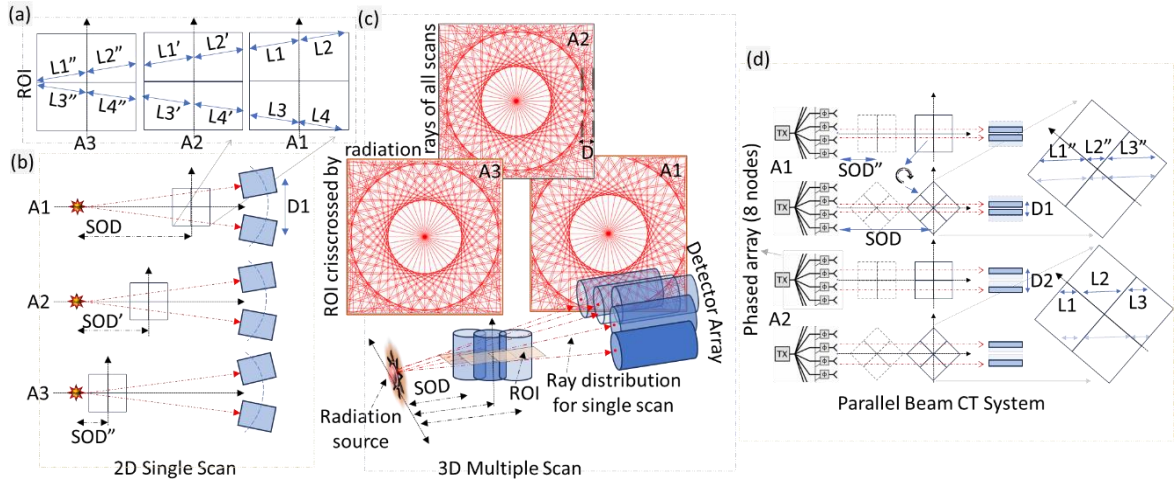

Figure 1: CT Scanning and effect of sample placement and inter-detector distance; (a) intercept length of radiation inside respective pixel for different sample locations, (b) three different CT configurations, (c) 3D representation of CT configuration and overlapped radiation for all projections highlighting the existence of sparse region, (d) parallel beam CT configurations showing sample location has no effect and inter-detector distance affects the intercept lengths of the weight matrix.

The estimation of intercept length (elements of weight matrix) is calculated using: (a) source to object distance (SOD), (b) fan angle of fan beam, (c) object/ROI size, and (d) detector array size. Typically, $W$ is a non-square sparse matrix, and specialized inverse algorithms are used to estimate back the $\lambda$ from measured $p$.

### 2.3. Optimality Parameters:

Two optimality criteria (based on the weight matrix) are proposed in this work and tested with respect to four CT reconstruction-based parameters. The flow chart to implement the algorithm is shown in figure 2.

### 2.3.1. Total Sum of Weight Matrix

Finding the inverse of a sparse matrix may or may not be an accurate mathematical operation and depends on the matrix's structure. It is shown in figs. 1(a) and 1(b) that depending on the sample location, the intercept length per pixel varies. Position A1 has a smaller intercept length, 'L2' as compared to position A2, i.e., L2'>L2. In fact, if we further slide the sample, placing it a little near the detector array will reduce this value L2≅0, undesirably adding to the sparsity of the matrix. Its corresponding L1 will also be relatively smaller. Physically, placing the sample on position A1 allows radiation to interact with sample material with a relatively longer mean-free path, thus imparting more information into measured data. The A1 location also allows equal interaction per pixel as L1<L2, but L1'≅L2' reduces the per-pixel biasedness. Similarly, if the sample is placed near the source, L1" may become relatively smaller. Figure 1(b) shows the crisscrossed overlap radiation paths for 36 views. Parameters D≅0 in A1 and A2 depict the space not covered after overall scanning. The example clearly explains that the sample location affects matrix structure. To estimate how many elements of the weight matrix contribute to this situation, as explained for L2≅0, one has to decide a threshold that seems not to be a very objective criterion. Simply finding the total sum of all elements in the matrix (Eq. 2) may account for L2≅0 and a relatively small L1 condition for this line integral and all of the line integrals. The object's size, source-to-object distance, fan beam angle or source-to-array to-array distance, and scanning array are the parameters required to estimate the elements of a typical weight matrix.

$$WM_{sum} = \sum_{\substack{0 < i \leq m \\ 0 < j \leq n}} W_{i,j} \quad (2)$$

For parallel beam CT systems, the $WM_{sum}$ will remain the same if SOD is varied. It will increase when the number of translations or detector array size is also increased. The former condition is used when only a single detector is used and translated to create a virtual array. The constraint to avoid ill-posedness enforces using the number of pixels $n$ = the number of translations. The usual $WM_{sum}$ thus fails to play as an optimal sample location estimation criterion alone and requires another criterion in support. However, if inter-detector distance $D1$ is varied by keeping the $n$ constant $WM_{sum}$ varies. This variation will generate $W^*$.

### 2.3.2. Condition Number ($k$):

An invertible matrix cannot have an eigenvalue equal to zero. The inverse of a matrix ($W$) with a high condition number, having eigenvalues near zero, will impart error in the estimation of $\lambda$ in eq. 1, if there is a slight error in the measurement of $p$. The condition number of matrix $W$ can be estimated by taking the ratio of maximum and minimum eigenvalues. In most cases, CT modeling results in a non-square large sparse or near sparse (with most elements near zero) matrix. Pre-conditioning methods such as principal component analysis (PCA) are a few of the classical approaches that improve the condition number of the matrix, stabilizing the inversion process and facilitating a relatively accurate solution. Hence, PCA is used for estimating $k$ if $W^*$ or $W$ is not invertible. Physically, placing the sample in an optimal location might result in a relatively smaller number of zeros or near-to-zeros matrix elements, thus improving the condition number and supporting the inverse problem solution towards accuracy.

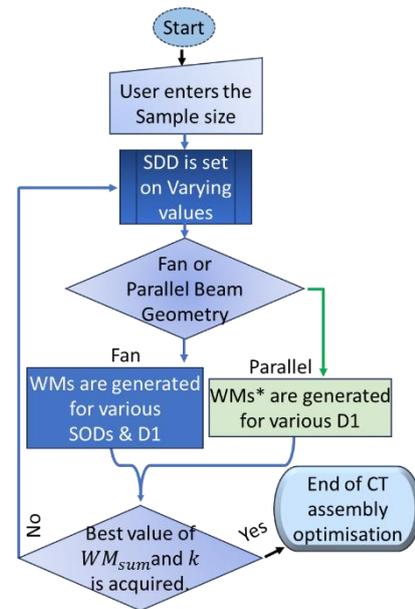

Figure 2: Flow diagram for CT assembly optimization before scanning.

### 2.4. Reconstruction-related Image metrics:

To test the performance of the above proposed CT assembly optimization criteria before scanning is done, the following reconstruction image-based performance parameters are used:

### 2.4.1. Standard Deviation of Sinogram:

A sinogram of a transmission CT is a collection of line integrals in the form of attenuated intensity or radiation counts. It contains information on the sample's inner profile, which is reconstructed in the form of an image by the reconstruction algorithm. If the sample has a homogeneous inner profile with no structure inside, the sinogram would appear plain and without any characteristics. However, for a given heterogeneous sample with some structure of finite boundary (for the sake of simplicity in discussion), if it is placed in three different locations, the location with the best coverage of the inner profile will contain most of the variation; thus projection data/sinogram will have a relatively large standard deviation. Suppose a sample is placed on a location where the radiation fails to cover most of the sample and touches either periphery or inner core with relatively less variation in profile. In that case, the radiation will interact with limited structures and may have relatively low standard variation in the sinogram.

### 2.4.2. KT-1

A scattering noise gets included in the sinogram, reducing the reconstructed image quality when radiation is not propagating perpendicular to the heterogeneous surfaces inside of the object. The optimal sample location will be the location that subjects the scanning of the sample to relatively low scattering noise. The goodness of fit obtained by applying Kanpur Thereom-1 gives an approximate estimate of scattering noise[5]. Root mean square error and Dice similarity coefficient are used as image metrics to compare the quality of reconstructed images concerning cyber replicas.

### 2.5. Experimental Setup:

Gamma CT system and sample details are given elsewhere. The source is Cs-137 of activity 1.5 µCi. The sample is made of Perspex embedded with an iron cylinder inside another aluminum cylinder at an off-centre location. It is shown in fig. Three on top left inset location. It is placed on three distances: (a) 18, 19.5, and 21 cm from the source. The three locations are also shown in fig 3. Five NaI (Tl) scintillator detectors are used to construct the scanning array. This array and source are kept fixed. The sample is rotated 36 times under 360 degrees with 10 degrees angle each.

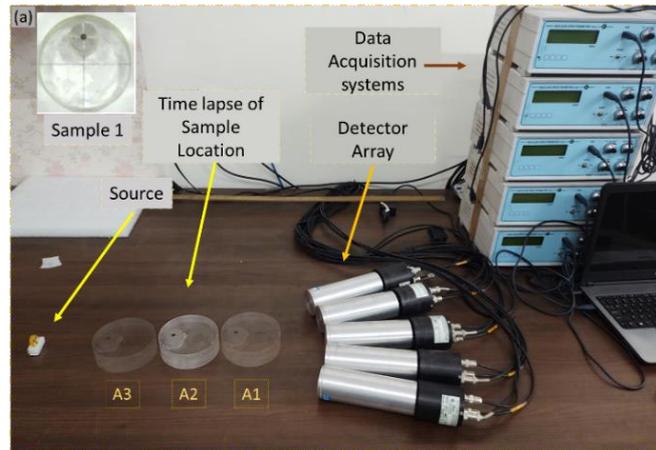

Figure 3. Gamma CT system and experiment

## 3. Results

The gamma CT data is collected for three different distances. The object and, thus, detectors are placed relatively near to the source, so the radiation wavefront spreads in a proper fan beam manner. With the distance, this radiation wave is expected to spread in the fan beam with a certain angle, thus increasing the probability of scatter inside the sample. However, the relative distance between the three sample locations is not large (~±7.5%). The simple inverse radon transform-based reconstruction algorithm is used. The goodness of fit of Kanpur Theorem -1 is estimated[5]. The cyber replica of sample 1 and reconstructed images for three different sample locations are shown in figs. 4(a), 4(b)m 4(c) and 4(d), respectively. Visually, fig. 4(c) matches with fig. 4(a). Corresponding weight matrices to three sample locations are used to estimate Condition numbers, the total sum of elements of these weight matrices. When estimating the minimum eigenvalues, components lesser than $\delta < 0.001$ are rejected[6]. Mask size equal to aluminum cylinder cross-section is used for dice coefficient estimation. The values are shown in fig. 5 in the bar plot. Sample location 2 has a slightly better condition number than the other two locations. The Total sum is relatively larger than other locations as well. The standard deviation of projection data for sample location A2 is relatively larger than for sample locations A1 and A3. A similar trend is shown after reconstruction by Dice, and root mean square values (RMSE) are estimated using normalized cyber replica as true image and goodness of fit of KT-1. The percentage value of RMSE is plotted with respect to RMSE for sample location 2.

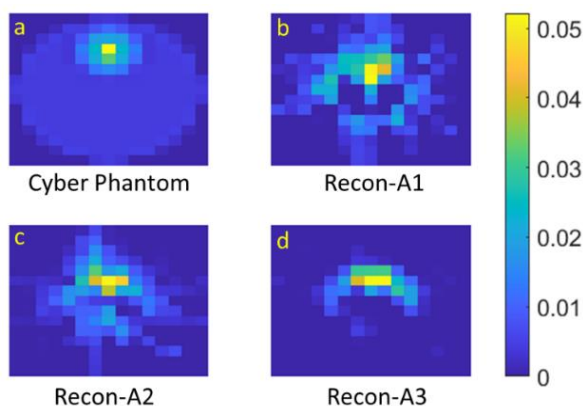

Figure 4: Reconstruction results for three different Sample Placement.

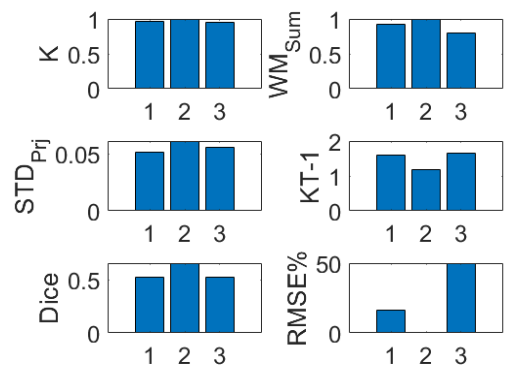

Figure 5: Optimal Sample Placement Strategy

## 4. Conclusion

The parameters obtained from the weight matrix, namely condition number and sum of elements, are shown to have a correlation with reconstruction image quality when CT assembly is optimized by varying associated structural parameters. These two parameters are affected by changing the size of the object, source-to-object distance, source-to-array distance, and inter-detector distance. Thus, it is recommended to optimize before scanning. The limitation of this implementation is that one may have to decide the value of $\delta$. This a-priori analysis is expected to help the CT system operator decide the scanning parameters, thus saving scanning time, the operational life of the source (in case an x-ray source is used), or any radiation dosage administered to a biological sample or patient.

## Reference

[1] Heinzl C, Kastner J, Amirkhamov A, Gröller E and Gusenbauer C 2012 Optimal specimen placement in cone beam X-ray computed tomography *NDT & E International* **50** 42–9

[2] Dabravolski A, Batenburg K J and Sijbers J 2014 Dynamic angle selection in X-ray computed tomography *Nucl Instrum Methods Phys Res B* **324** 17–24

[3] Goswami M, Shakya S, Saxena A and Munshi P 2015 Reliable reconstruction strategy with higher grid resolution for limited data tomography *NDT & E International* **72** 17–24

[4] Kak A C and Slaney M 2001 Principles of Computerized Tomographic Imaging *Principles of Computerized Tomographic Imaging*

[5] Kumari K and Goswami M 2023 Relative estimation of scattering noise and electronic noise of a radiation detector *Meas Sci Technol* **34** 125403

[6] Stewart G W 2001 A Krylov-Schur Algorithm for Large Eigenproblems *SIAM J. Matrix Anal. Appl.* **23** 601–14